\documentstyle[prl,aps,rotate,array,multicol,epsf]{revtex}
\newcommand{\lrule}{ \end{multicols} \noindent
  \rule{0.5\textwidth}{0.1mm}\rule{0.1mm}{3pt}\newline }
\newcommand{\rrule}{ \noindent \parbox{\textwidth}{
  \hfill\rule[-3pt]{0.1mm}{3pt}\rule{0.5\textwidth}{0.1mm}}
  \begin{multicols}{2}\noindent }

\begin{document}

\draft

% ***************************** TITLE **********************************

\title{Interplay of spin-discriminated Andreev bound states forming the
$0$-$\pi$ transition in Superconductor-Ferromagnet-Superconductor
Junctions}
% ************************ AUTHOR & ADDRESS ****************************
\author{Yu.~S.~Barash$^{1,2}$, I.~V.~Bobkova$^2$}
\address{$^1$Center for Electronic Correlations and Magnetism,
Institute of Physics, University of Augsburg, D-86135 Augsburg, Germany\\
$^2$ Lebedev Physical Institute, Leninsky Prospect 53,
 Moscow 119991, Russia\\
}

% ****************************** DATE **********************************

%\date{\today}

% *************************** MAKE TITLE *******************************

\maketitle

% **************************** ABSTRACT ********************************

\begin{abstract}
The Josephson current in {\it S-F-S} junctions is described by
taking into account different reflection (transmission) amplitudes
for quasiparticles with spin up and down. We show that the
$0$-$\pi$ transition in the junctions can take place at some
temperature only for sufficiently strong spin-activity of the
interface. In particular, Andreev interface bound state energies
in one spin channel have to be all negative, while in the other
one positive. Only one spin channel contributes then to the
zero-temperature Josephson current. At the temperature of the
$0$-$\pi$ transition two spin channels substantially compensate
each other and can result in a pronounced minimum in the critical
current in tunnel junctions. The minimal critical current is
quadratic in small transparency and contains first and second
harmonics of one and the same order.
\end{abstract}

% ************************** PACS NUMBERS ******************************

%\pacs{PACS numbers: 74.50.+r, 74.80.Fp}

% TIMESTAMP
%\centerline{\today}

% ************************** BODY OF PAPER *****************************
\begin{multicols}{2}
Growing interest at the time being to spin-dependent transport in
superconductor-ferromagnet systems concerns, in particular, the dc
Josephson effect. The possibility for forming the $\pi$-junction
owing to interfaces containing magnetic impurities, was indicated
for the first time in\cite{bul77}. Since then the supercurrent
across magnetically active interfaces has been studying
theoretically both in {\it S-FM-S} junctions with a ferromagnetic
metal separating
superconductors\cite{buz82,buz92,demler97,heik00,koshina01,volkov01},
and in {\it S-FI-S} junctions with interfaces made of a
ferromagnetic insulator or
semiconductor\cite{mrs88,fogel00,fogel01}. The $0$-$\pi$
transition in {\it S-F-S} junctions was predicted under certain
conditions in both cases\cite{buz82,buz92,fogel00}. Recently the
effect has been observed experimentally in {\it S-FM-S}
junctions\cite{ryazanov01}.

$0$-$\pi$ transition in {\it S-FM-S} highly transparent junctions
is mostly discussed with respect to the proximity effect in a
ferromagnetic metal\cite{buz82}. An exchange field in a
ferromagnetic metal between two superconductors induces specific
oscillations in the exponentially decaying Cooper pair density.
For this reason an exchange field dependent oscillations in the
critical current can arise in {\it S-FM-S} sandwiches. In {\it
S-FI-S} junctions the proximity effect in a ferromagnetic
insulator or semiconductor usually is much weaker, as compared
with the case of a ferromagnetic metal, and can be disregarded. It
has been found experimentally that a ferromagnetic semiconductor
represents a ferromagnetic barrier in tunnel junctions, providing
different transmission probabilities for up and down
spins\cite{tedrow86,hao90}. In {\it S-FM-S} tunnel junctions the
proximity in a ferromagnetic metal does not manifest itself in the
critical current in the dominating, linear in small transparency
term. In all these cases spin-discrimination by the interface is
especially important, as it results in proximity effects induced
by a ferromagnetic layer in adjacent superconducting regions. If
the interface thickness is less than the superconducting coherence
length, the interface effects on the junction properties are
conveniently described by the $S$-matrix approach. As this was
demonstrated in\cite{fogel00}, a quasiparticle scattering on
magnetically active interfaces can themselves lead to a formation
of a $\pi$-junction, even in the absence of any proximity-induced
processes inside interfaces. The physics for this is associated
with interface Andreev bound states caused by the
spin-discriminating processes (for instance, by the effects of an
exchange field in a ferromagnet). In the present paper we report
the spectra of Andreev interface states, which arise on
magnetically-active interfaces with arbitrary spin-dependent
reflection and transmission amplitudes, and analyze their
interplay in forming the Josephson current.

Since interface bound states depend explicitly on the phase
difference, they are directly associated with the Josephson
current through the junction. This is in accordance with the
general relation between the Josephson current and the spectra of
Andreev interface bound states\cite{been291}. We develop on this
basis comparatively simple analytical description of the dc
Josephson current in {\it S-F-S} junctions, when the proximity in
ferromagnets is not important. We find exact conditions for the
presence of the $0$-$\pi$ transition and demonstrate that
interface bound states from different spin channels have to be
strongly discriminated in this case. In particular, both interface
bound state energies in one spin channel have to be negative,
while positive in the other channel.

Our approach is based on the quasiclassical formulation of the
superconductivity\cite{eil68,larkin68,larkin75,el71,larkin86,rainer83}.
The quasiclassical equations, as is known, have to be completed
with respective boundary conditions. For smooth flat surfaces or
interfaces not distinguishing quasiparticle spin directions, the
boundary conditions for the quasiclassical Green's function were
derived for the first time in\cite{zaitsev84}. Later on they were
generalized to incorporate spin-active potentials\cite{mrs88}.
This permitted to solve some particular problems for systems with
impenetrable spin-active boundaries\cite{tsr88,kulic00} and for
junctions in the tunneling limit\cite{mrs88}. The possibility for
studying various magnetically active interfaces with finite
transmission has appeared only recently, when the particular
formulation of the quasiclassical theory with its basic
quantities, equations, the corresponding boundary and asymptotic
conditions, was substantially modified and simplified. The
achievements are associated with making clear a general structure
of the quasiclassical matrix Green's function, reducing the
calculation of the Green's function to finding its ingredients,
the so-called Riccati amplitudes or coherence
functions\cite{schopohl98,eschrig00,fogel00}. For magnetically
active interfaces this formulation was developed in\cite{fogel00}.
Using this approach, we present analytical results, explicitly
describing how magnetically-active interfaces with spin-dependent
transmission amplitudes influence the Josephson current.

Consider a smooth plane interface between two superconductors or
normal metals. Interface is characterized by the normal-state
scattering $\cal S$ matrix, which can be described as follows.
Exploiting Pauli-matrices $\hat\tau_j$ in particle-hole space, a
scattering matrix is represented as ${\cal
S}=S(1+\hat\tau_z)/2+\widetilde S(1-\hat\tau_z)/2$, where
$\widetilde S(p_\|)=S^{tr}(-p_\|)$. Each component $\hat{S}_{ij}$
in matrix $S=\|\hat{S}_{ij}\|$ ($i(j)=1,2$) is in its turn a
matrix in spin space. Matrix $\hat{S}_{ii}$ contains, in general,
spin-dependent reflection amplitudes of normal-state
quasiparticles from the interface in $i$-th half-space, while
$\hat{S}_{ij}$ with $i\ne j$ incorporates spin-dependent
transmission amplitudes of normal-state quasiparticles from side
$i$. For the interface potentials conserving particle current, the
scattering matrix has to satisfy the unitarity condition: ${\cal
S}{\cal S}^\dagger=1$. If the interface Hamiltonian possesses the
time-reversal symmetry, one gets an additional constraint on the
scattering matrix: $ S(\bbox{p}_f, \bbox{\mu})=\hat \sigma_y
S^{tr}(-\underline{\bbox{p}}_f,-\bbox{\mu})\hat \sigma_y
$\cite{mrs88}. Assuming the scattering matrix diagonal in spin
space and obeying $S(p_\|)=S(-p_\|)$, one obtains from here $ \hat
S_{12}(\bbox{p}_f)=\hat S_{21}(\bbox{p}_f)=\hat d(\bbox{p}_f)$.
For a barrier potential of the form ${\hat
V(x)}=V(x)\hat{1}+{\bbox\mu(x)} \hat{\bbox\sigma}$, it is
convenient to take the $z$ axis along the only characteristic
"magnetization" vector ${\bbox\mu}$. Then $\hat{S}_{ij}$-matrices
are diagonal. Diagonal components of $ \hat{S}_{11}$  and
$\hat{S}_{22}$ are $r_{1,\uparrow(\downarrow)}$ and
$r_{2,\uparrow(\downarrow)}$ respectively, and the diagonal
components of $\hat{S}_{12}=\hat{S}_{21}$ are
$d_{\uparrow(\downarrow)}$. On account of the above relations one
also obtains $\widetilde S(\bbox{p}_f)=S(\bbox{p}_f)$.

It follows from the unitarity of the scattering matrix
$r_{2,\uparrow(\downarrow)}d^*_{\uparrow(\downarrow)}+d_{\uparrow(\downarrow)}
r^*_{1,\uparrow(\downarrow)}=0$,
$|r_{1,\uparrow(\downarrow)}|^2+|d_{
\uparrow(\downarrow)}|^2=|r_{2,\uparrow(\downarrow)}|^2+|d_{
\uparrow(\downarrow)}|^2=1$. Remembering this and introducing
$r_{1,\uparrow(\downarrow)}=|r_{\uparrow(\downarrow)}|
e^{\Theta_{1,\uparrow(\downarrow)}}$,
$r_{2,\uparrow(\downarrow)}=|r_{\uparrow(\downarrow)}|
e^{\Theta_{2,\uparrow(\downarrow)}}$, we get
$
d_{\uparrow}d_{\downarrow}=\alpha|d_{\uparrow}||d_{\downarrow}|\exp\left(
\frac{\displaystyle i}{\displaystyle 2}(\Theta_{1,\uparrow}+
\Theta_{1,\downarrow}+\Theta_{2,\uparrow}+\Theta_{2,\downarrow})\right)
\enspace . \label{rd}
$
Here $\alpha=\pm1$. One can show that $\alpha=-1$ for a
nonmagnetic barrier, while $\alpha=1$ for a
purely magnetic ($V=0$) and sufficiently high barrier.
For a rectangular potential
$\alpha=-{\rm sgn}[(V-(p^2_{f,x}/2)+h)(V-(p^2_{f,x}/2)-h)]$, where
$h=\mu_z$. Hence, $\alpha=1$, if the wave function in one spin channel
exponentially decays in the barrier region, while in the other
channel it oscillates. In general, $\alpha$ can depend on quasiparticle
momentum direction.

We have calculated {\it spectra of interface bound states on
magnetically active interfaces} with the scattering $\cal
S$-matrix described above. Our main results are as follows.

For a symmetric barrier potential $\hat{V}(-x)=\hat{V}(x)$, when
$\Theta_{1,\uparrow(\downarrow)}=\Theta_{2,\uparrow(\downarrow)}$,
two branches of energies of the Andreev interface
bound states in {\it S-F-S} symmetric junctions, corresponding
to one spin channel (spin up for electron-like quasiparticles), take the form
\begin{equation}
\varepsilon_{\pm}=|\Delta|\ {\rm sgn}\left(\sin
\frac{\Phi_{\pm}}{2}\right)\cos \frac{\Phi_{\pm}}{2} \enspace ,
\label{cd}
\end{equation}
where
\begin{equation}
\Phi_{\pm}(\alpha,\chi)=\Theta \pm {\rm arccos}\left[
\sqrt{R_\uparrow R_\downarrow}-\alpha \sqrt{D_\uparrow
D_\downarrow}\cos \chi \right]  , \label{phipm}
\end{equation}
$R_\uparrow (\downarrow)=|r_{\uparrow(\downarrow)}|^2$,
$D_\uparrow (\downarrow)=|d_{\uparrow(\downarrow)}|^2$ and
$\Theta=\Theta_{\uparrow}-\Theta_{\downarrow}$. The solution for
the other spin channel is obtained from Eq.(\ref{phipm}) by the
substitution $\Theta\to -\Theta$. Energies $\varepsilon_{\pm}$
implicitly depend on quasiparticle momentum directions via the
parameter $\Theta$, reflection and transmission coefficients, and,
possibly, $\alpha$.

Eq.(\ref{phipm}) describes, in particular, how spin-filtering
effects suppress the Josephson current, when the transmission
coefficient for quasiparticles with one spin orientation is
sufficiently small as compared to the other one. As it is seen,
spin-dependent transmission (and reflection) coefficients enter
the spectra of interface Andreev bound states (as well as the
Josephson current) as an effective transparency
$\sqrt{D_{\uparrow}D_{\downarrow}}$ (and a reflectivity
$\sqrt{R_{\uparrow}R_{\downarrow}}$). There is, however, no
general prescriptions for replacing spin-independent coefficients
by spin-dependent ones without particular calculations, since the
relations $R_{\uparrow(\downarrow)}=1-D_{\uparrow(\downarrow)}$
make it ambiguous. According to Eqs.(\ref{cd}), (\ref{phipm}), the
difference between spin-dependent phases of reflection amplitudes
$\Theta=\Theta_{\uparrow}-\Theta_{\downarrow}$ plays a crucial
role, lifting spin-degeneracy of the Andreev bound states. In the
particular case $D_\uparrow=D_\downarrow$, $\alpha=-1$ the whole
bound state spectra of two spin channels reduce to those found in
\onlinecite{fogel01}. For an impenetrable spin-active surface the
spectrum Eq.(\ref{cd}) transforms to $\varepsilon_{B,0}=|\Delta|\
{\rm sgn}\left(\sin \frac{\Theta}{2}\right)\cos \frac{\Theta}{2}$.
Then the whole spectra of two spin channels
$\varepsilon_{B,0}=\pm|\Delta|\cos \frac{\Theta}{2}$ coincide with
obtained in\cite{fogel00}. For a nonmagnetic interface
($D_\uparrow=D_\downarrow=D$, $\Theta=0$, $\alpha=-1$) our result
Eq.(\ref{cd}) leads to well known one positive and one negative
spin-degenerated interface Andreev bound
states\cite{fur90,fur191,been191}
$\varepsilon_B=\pm|\Delta|\sqrt{1-D\sin^2(\chi/2)}$. With
increasing parameter $\Theta$ bound states in Eq.(\ref{cd}) can
change their signs both continuously or abruptly, due to a factor
${\rm sgn}\left(\sin\left(\Phi_\pm/2\right)\right)$ in the latter
case. For this reason, under certain conditions, both levels in
one spin channel can be positive (or negative) at the same time.
Energy spectrum formed jointly by two spin channels is symmetric
with respect to the sign change.

Generic form of the spectrum given by Eq.(\ref{cd}) always arise,
if order parameters for the incoming and the outgoing
quasiparticle momenta in a reflection (or transmission) event
differ, in fact, only by their phases, irrelevant to a physical
origin of the phase difference $\Phi$. This applies, in
particular, to Andreev surface states on nonmagnetic impenetrable
boundary, in the presence of the phase difference. Bound state
spectrum Eq.(\ref{cd}) has been obtained assuming spatially
constant order parameter. Effects of the self-consistency on the
results are not crucial. For instance, in the low-energy range
(e.i. for $\Phi$ close to $\pi$) the spectrum (\ref{cd}) remains
to be valid for spatially dependent $|\Delta(x)|$ as well, if one
substitutes $|\Delta|$ for effective surface order parameter
$|\tilde\Delta(0)|$, defined in\cite{barash00}.

Consider now {\it the Josephson current in a symmetric $S-F-S$
junction}. The Josephson current is flowing via the bound states
(\ref{cd}), analogously to what takes place in nonmagnetic
symmetric junctions\cite{fur90,fur191,been191,been291}. Hence, in
a quantum point contact with spin-active constriction the total
Josephson current carrying by two spin channels can be found as $
J(\chi,T)=2e\sum \limits_m \frac{\displaystyle d
\varepsilon_m}{\displaystyle d \chi}n(\varepsilon_m)$ $= -2 e \sum
\limits_{\varepsilon_m>0} \frac{\displaystyle d
\varepsilon_m}{\displaystyle d \chi}{\rm tanh}\frac{\displaystyle
\varepsilon_m}{\displaystyle 2 T}$\enspace . With Eqs.
(\ref{cd}),(\ref{phipm}) we find \lrule
\begin{equation}
J(\chi,T)= A(\chi)\left[ \sin \left(\frac{\displaystyle
\Phi_{+}}{\displaystyle 2}\right){\rm
tanh}\left(\frac{\displaystyle |\Delta| \cos
\frac{\Phi_{+}}{2}}{\displaystyle 2 T}\right)-\sin
\left(\frac{\displaystyle \Phi_{-}}{\displaystyle 2}\right){\rm
tanh}\left(\frac{\displaystyle |\Delta| \cos \frac{ \Phi_{-}}{
2}}{\displaystyle 2 T}\right) \right] \enspace , \label{qpc}
\end{equation}
where $A(\chi)= -\alpha e |\Delta| \sqrt{D_\uparrow
D_\downarrow}\sin \chi\left[1-\left( \sqrt{R_\uparrow
R_\downarrow}-\alpha \sqrt{D_\uparrow D_\downarrow}\cos \chi
\right)^2\right]^{-1/2}$.\\ \rrule In the absence of a
spin-activity $\bf{\mu}(x)=0$ and, hence, $\Theta=0$, $\alpha=-1$,
$D_{\uparrow}=D_{\downarrow}$. Then Eq.(\ref{qpc}) reduces to
well-known contributions from two spin-degenerated channels to the
Josephson current of quantum point contact with finite
transparency. Spin-discrimination by the interface lifts the
degeneracy. The current flowing in the spin-up channel takes the
form $J_\uparrow(\chi,T)=-A(\chi)\left(\left|\sin\left(
\frac{\displaystyle\Phi_+}{\displaystyle2}\right)\right|n_f\left(\frac{\displaystyle
\varepsilon_+}{\displaystyle T}\right)-\left|\sin\left(\frac{
\displaystyle\Phi_-}{\displaystyle2}\right)\right|n_f\left(\frac{
\displaystyle\varepsilon_-}{\displaystyle T}\right)\right)$.
$J_\downarrow(\chi,T)$ is obtained from here by the interchange
$\Theta\rightarrow -\Theta$.

\narrowtext
\begin{figure}[bt]
\epsfxsize=8.5cm\hfil\epsfbox{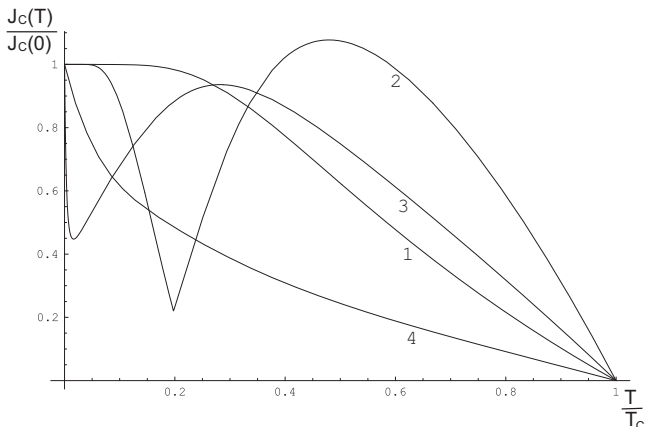} \hfill\caption[bt]{
Critical current $J_c(T)$, normalized to its value at zero
temperature $J_c(0)$ and taken for various values of $\Theta$:
$\Theta=0.4\pi$ (1), $\Theta=0.7\pi$ (2), $\Theta=0.8\pi$ (3),
$\Theta=0.9\pi$ (4). Transparencies are
$D_{\uparrow}=D_{\downarrow}=0.1$ and $\alpha=-1$. }
\label{fig1a.eps}
\end{figure}

Andreev bound states in different spin channels can carry current
in opposite directions, so that one direction prevails in the
total current at low temperatures, while the other one near $T_c$.
We find that in the case $\sqrt{R_\uparrow R_\downarrow}-\alpha
\sqrt{D_\uparrow D_\downarrow}\cos\chi>0$ the current
Eq.(\ref{qpc}) changes its sign with varying the temperature at a
given phase difference $\chi$, if
$\pi/2<|\Theta|<\pi-\arccos\left[\sqrt{R_\uparrow
R_\downarrow}-\alpha \sqrt{D_\uparrow D_\downarrow}\cos
\chi\right]$. Analogously, for $\sqrt{R_\uparrow
R_\downarrow}-\alpha \sqrt{D_\uparrow D_\downarrow}\cos\chi<0$ the
condition for a sign change is $\pi-\arccos\left(\sqrt{R_\uparrow
R_\downarrow}-\alpha \sqrt{D_\uparrow D_\downarrow}\cos
\chi\right)<|\Theta|<\pi/2$. The above conditions imply, in
particular, that both interface Andreev states in one spin channel
have positive energies while the energies in the other channel are
negative.

\begin{figure}
\centerline{{\epsfxsize=0.8\textwidth{\epsfbox{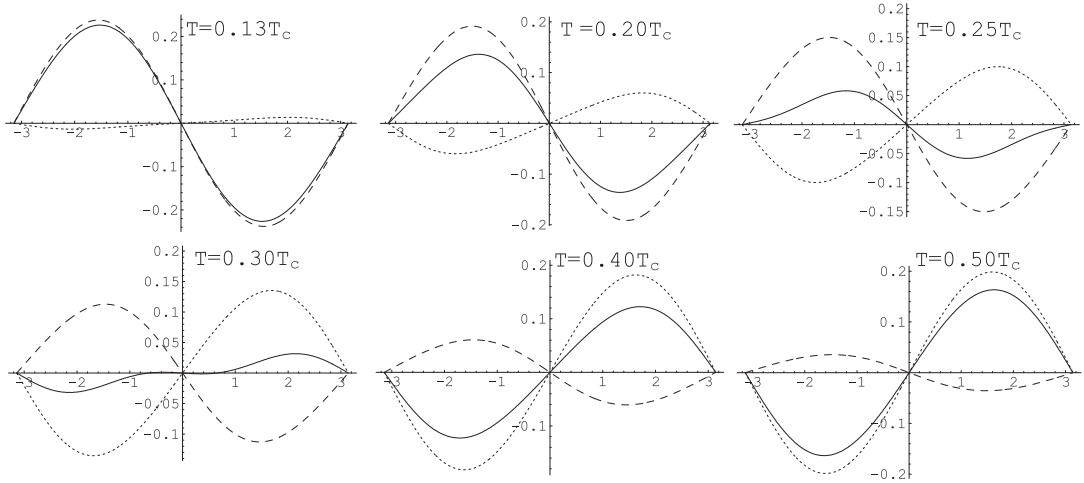}}}}
\caption[]{Current-phase relations at various temperatures for the
total Josephson current (solid line) and separate contributions of
spin-up(dashed line) and spin-down (dotted line)channels. The
parameters are chosen to be $D_\uparrow=D_\downarrow=0.05$,
$\Theta=2\pi/3$. The current is normalized to the zero-temperature
critical current in the nonmagnetic case $\Theta=0$.}
\label{fig2.ps}
\end{figure}

In tunnel junctions, where the former condition holds, the
interplay of two spin channels, in a certain rage of $\Theta$,
results in a pronounced minimum of the Josephson critical current.
This is shown in Fig.\ \ref{fig1a.eps}. The appearance of the
minimum can be explained as follows. Since at zero temperature
only quasiparticle states with negative energies are occupied,
only the respective spin channel contribute the zero-temperature
Josephson current. The contribution from the second channel rises
with increasing temperature and becomes important at temperatures
of the order of positive bound state energies. The corresponding
current in the second channel turns out to be aligned in the
opposite direction as compared to the zero-temperature current.
Competing contributions of different spin channels result in the
$0$-$\pi$ transition in the junction. Near the transition currents
from two channels substantially compensate each other. The
interplay of two spin channels forming the Josephson current, is
shown at various temperatures in Fig.\ \ref{fig2.ps}. $0$-$\pi$
transition takes plays at the temperature, where maximal currents
in ``$0$'' and ``$\pi$'' spin channels become equal to each other.

\end{multicols}
\widetext

\begin{multicols}{2}
As this was indicated in\cite{fogel00}, the $0$-$\pi$ transition
takes place abruptly. It is instructive to analyze this problem in
the case $D_{\uparrow(\downarrow)}\ll 1$. In accordance with
Eq.(\ref{qpc}), the anomalous temperature behavior of the
Josephson critical current in $S-F-S$ junctions with small
transparencies takes the form \lrule
\begin{equation}
J_c(T)=-\alpha e |\Delta|\sqrt{D_\uparrow D_\downarrow} \left[
\frac{\displaystyle \varepsilon_{B,0}(\Theta)}{\displaystyle
|\Delta|}{\rm tanh}\left(\frac{\displaystyle
\varepsilon_{B,0}(\Theta)}{\displaystyle 2
T}\right)-\frac{\displaystyle |\Delta|}{\displaystyle 2
T}\frac{\displaystyle 1 - \frac{\displaystyle
\varepsilon^2_{B,0}(\Theta)}{\displaystyle
|\Delta|^2}}{\displaystyle \cosh^2
\left(\frac{\varepsilon_{B,0}(\Theta)}{2 T}\right)} \right]
\enspace , \label{tunneltok}
\end{equation}
\rrule
where $\varepsilon_{B,0}(\Theta)=|\Delta\cos
\frac{\Theta}{2}|$ is the positive bound state energy on an
impenetrable spin-active surface.

In accordance with Eq.(\ref{tunneltok}), a part of the critical
current, which is linear in effective transmission coefficient
$\sqrt{D_{\uparrow}D_{\downarrow}}$, changes its sign going
through zero at some temperature. The tunnelling contributions
from two spin channels cancel each other there. Quadratic in
transparency terms, however, survive and determines the
current-phase relation and the critical current itself in the
vicinity of the transition temperature. One can get from
Eq.(\ref{qpc}) and see from Fig.\ \ref{fig2.ps}, that quadratic in
transmission first and second harmonics have one and the same
order of value.  At the temperature, where the linear in
transmission term vanishes (near the transition temperature) the
critical current $\propto D_\uparrow
D_\downarrow\left(\sin\chi+\frac{\displaystyle 1}{\displaystyle
2}\sin2\chi\right)$. Thus, on account of quadratic in transmission
terms, the total critical current does not vanish when the
transition from $0$ to $\pi$ junction takes place. The minimal
value of the critical current $\propto D_{\uparrow}D_{\downarrow}$
if the transparencies are small.

The second harmonic in the Josephson current Eq.(\ref{qpc}) is
also important for junctions with sufficiently high transparency.
For fully transparent junctions with $\Theta=\pi/2$ the
current-phase relation has a period $\pi$, which leads, in
particular, to a half-periodicity of the dependence of the
critical current on an applied magnetic field.

For $\Theta=\pi-\delta$($|\delta|\ll1$)we find from
Eq.(\ref{qpc}), that the current $\propto 1/T$ under the condition
$|\Delta||\delta|/4\ll T\ll \frac{\displaystyle
|\Delta|}{\displaystyle 4}
\sqrt{D_{\uparrow}+D_{\downarrow}+2\alpha\sqrt{D_{\uparrow}D_{\downarrow}}\cos\chi}$.
 This anomalous temperature behavior is
associated with the presence of zero-energy (or low-energy)
surface bound states in both banks of the junction with vanishing
transparency(see also\cite{fogel00}). It quickly disappears with
increasing transparency, when only one energy of two different
bound states in the spin channel can be close or even equal to
zero, while the other is roughly the order of $\Delta$ (see
Eqs.(\ref{phipm}), (\ref{cd})).

In the case of classical junctions one should carry out the
integration over the Fermi surface in the presence of momentum
dependent bound state energies. This does not modify strongly our
main results, obtained above for quantum point contacts. Although,
$1/T$ behavior can be substantially distorted or even smeared out
in this case. For instance, if only a maximum (or a minimum) of
dispersive bound states is equal to zero in {\it S-F-S} tunnel
junctions with two-dimensional $s$-wave superconductors, the
current in a classical junction would be $\propto 1/\sqrt{T}$.

In conclusion, we have studied the interplay of spin-discriminated
channels on the spectrum of Andreev bound states and the Josephson
current through a ferromagnetic interface. The conditions for the
$0$-$\pi$ transition in the junction are found. They imply strong
discrimination of Andreev interface states in two spin channels.
In particular, only one spin channel contribute to the
zero-temperature Josephson current in this case. Linear in
transparency Josephson current vanishes near the $0$ - $\pi$
transition. The critical current in its minimum is quadratic in
small transparency and contains two first harmonics $\sin\chi$ and
$\sin2\chi$ which are of the same order there.

{\it Acknowledgments} Yu.B. would like to thank T. Kopp and H.
Kroha for useful discussions of the problem considered above and
L. Barash for a technical help. This work was supported in part by
BMBF 13N6918/1 (Yu.B.) and by the Russian Foundation for Basic
Research under Grant No. 99-02-17906 (Yu.B. and I.B.)

% ************************** REFERENCES ********************************

% **************************** FIGURES *********************************

%\newpage

% **************************** TABLES **********************************
\end{multicols}
\end{document}